\documentclass[apjl]{emulateapj}
\renewcommand\abstractname{\textbf{ABSTRACT}}

\usepackage{amssymb}
\usepackage{booktabs}
\usepackage{graphicx}
\usepackage{pbox}

\renewcommand{\phi}{\varphi}

\usepackage{ amssymb }

\begin{document}

{
\title{The Feasibility of Using Black Widow Pulsars in Pulsar Timing Arrays for Gravitational Wave Detection}
\author{Christopher Bochenek$^{1,2}$, Scott Ransom$^{1}$, Paul Demorest$^3$}
\affiliation{1. National Radio Astronomy Observatory, Charlottesville, VA \\ 2. Physics Department, University of Chicago, Chicago, IL \\ 3. National Radio Astronomy Observatory, Socorro, NM}
\begin{abstract}
     In the past five years, approximately one third of the 65 pulsars discovered by radio observations of Fermi unassociated sources are black widow pulsars (BWPs). BWPs are binary millisecond pulsars with companion masses ranging from 0.01-0.1 solar masses which often exhibit radio eclipses. The bloated companions in BWP systems exert small torques on the system causing the orbit to change on small but measurable time scales. Because adding parameters to a timing model reduces sensitivity to a gravitational wave (GW) signal, the need to fit many orbital frequency derivatives to the timing data is potentially problematic for using BWPs to detect GWs with pulsar timing arrays. Using simulated data with up to four orbital frequency derivatives, we show that fitting for orbital frequency derivatives absorbs less than 5\% of the low frequency spectrum expected from a stochastic gravitational wave background signal. Furthermore, this result does not change with orbital period.  Therefore, we suggest that if timing systematics can be accounted for by modeling orbital frequency derivatives and is not caused by spin frequency noise, pulsar timing array experiments should include BWPs in their arrays.
\end{abstract}
\section{Introduction}
\par
Black widow pulsars (BWPs) are a special class of binary millisecond pulsars (MSPs). They often have orbital periods ranging from 2-20 hours \citep{Chenetal2013}. However, the defining feature of BWPs is their low mass companions, with masses much less than $0.1$ M$_{\odot}$ \citep{Roberts2013}.
\par
Even though BWPs have very small companions, the ionized gas from the companions often eclipses the pulsar \citep{Robertsetal2014}. This occurs because the pulsar wind blows stellar material away from the companion, but leaves plasma in the system, creating a screen for the pulsar's radio emission \citep{Thompson1995}. The eclipse in the first discovered black widow system, PSR B1957+20, covers about $10\%$ of the orbit \citep{Fruchteretal1988, Robertsetal2014}.
\par
However, in this work, we are most concerned with the stochastic variation in orbital period exhibited by BWP systems. In fact, in a few years of observing PSR B1957+20, the sign of the orbital period derivative changed \citep{Arzoumanianetal1994}. In order to explain this observation, \citet{Applegateetal1994} gave a mechanism for the changes in orbital period over time. Their model starts from the fact that there are large tidal forces between the pulsar and companion due to their tight orbit. These tidal forces drive stellar convection in the companion. Combined with rapid rotation, stellar convection gives rise to a strong magnetic field. This strong magnetic field and differential rotation in the companion cause the companion to become more or less oblate \citep{Applegate1992}. If the companion becomes more oblate, the centrepital force between the pulsar and the companion increases while their angular momentum remains constant, causing a decrease in orbital period. If the companion becomes less oblate, the centrepital force decreases causing the orbital period to increase.
\par
Due to the high orbital variability of BWP systems, precisely modeling the pulse arrival times for the pulsar often requires modeling the orbit with many orbital frequency derivatives (OFDs) in our timing solutions. However, fitting many OFDs reduces sensitivity to gravitational waves (GWs). Because the parameters of a pulsar are determined from the same data used to search for GWs, some of the GW signal will be absorbed in fits for pulsar timing parameters, as the GW signal will appear in at least a small way to be covariant with the fit parameters \citep{Demorestetal2013}. This process reduces the amount of GW signal in the data after the timing fit, reducing sensitivity to a GW source.
\par
\citet{Niceetal2000} showed PSR B1957+20, the first discovered BWP system, had what appeared to be significant stochastic deviation from the best fit timing model, or timing noise, over many years and suggested that the source of this noise is intrinsic to the system, such as stochastic variation in spin period. This result has led pulsar timing array (PTA) collaborations to be reluctant about including BWPs in PTAs to search for GWs. The variations in orbital period make BWPs difficult to time and perhaps contribute to the timing noise. However, if we can account for the timing variations in a way that does not significantly reduce sensitivity to GWs, then black widow pulsars should be considered for inclusion in PTAs. We show that modeling orbital frequency derivatives does not significantly reduce sensitivity to GWs and suggest that other timing variations can be explained by variations in dispersion measure.
\par 
All of this comes at a time when BWPs are being discovered in droves. One of the most successful ways of searching for MSPs has been surveying $\gamma$-ray sources detected by the Fermi-LAT \citep{Rayetal2012,Roberts2013}. Among these newly discovered Fermi MSPs, about a third of the 65 pulsars discovered using this method are BWPs \citep{Rayetal2012,Roberts2013}. Given that BWPs are being discovered at a very high rate, it makes sense to revisit the question of whether BWPs can be timed precisely enough to use in PTAs. 

\section{Methods}
\subsection{Quantifying the Amount of Gravitational Wave Signal Post-fit}
\par
When a GW passes between the Earth and a pulsar, the distance between the pulsar and the Earth is altered as a function of time \citep{Detweiler1979}. Therefore, there will be timing variations from GWs. \citet{Jaffeetal2003} showed that the characteristic strain spectrum for a stochastic background of supermassive black hole binaries will be a power law of index -2/3. Thus, we can estimate the effect of such a background on a pulsar's times of arrival (TOA). By hypothesizing that all of the structure in the timing residuals is due to gravitational waves, we can place a limit on the amplitude of this background \citep[e.g.][]{Kaspietal1994}.
\par
In order to quantify how much gravitational wave signal is absorbed by fitting orbital frequency derivatives, we use a method from \citet{Demorestetal2013} that assumes there is a stochastic gravitational wave background signal of spectral index -2/3 in a data set and calculates how much of that assumed signal will be present in the post-fit timing residuals. First, we calculate the pre-fit gravitational wave covariance matrix, {\bf C}$_{{\bf y}}^{{\bf GW}}$ \citep[see][]{Demorestetal2013,vanHaasterenetal2009}. This is done by calculating the statistical expectation value of $y_ a(t_i)y_a(t_j)$ for all pulse TOAs $i$ and $j$, where $y_a(t)$ is the amount a gravitational wave background signal shifts a pulse TOA. Then, we apply the timing fit to this matrix to get the post-fit, or residual, gravitational wave covariance matrix, {\bf C}$_{{\bf r}}^{{\bf GW}}$. Next, we weight {\bf C}$_{{\bf r}}^{{\bf GW}}$ by the diagonal matrix of pulse TOA uncertainties, {\bf W}, to get {\bf W}{\bf C}$_{{\bf r}}^{{\bf GW}}${\bf W} . Then, we diagonalize {\bf W}{\bf C}$_{{\bf r}}^{{\bf GW}}${\bf W}  to form an orthonormal basis of eigenvectors that is completely orthogonal to the timing fit, optimally capturing the gravitational wave signal left in the residuals after the fit. 
\par
The eigenvalues $\lambda_i$ are the square of the amplitude each element of this basis contributes to the timing residuals. By summing over $\sqrt{\lambda_i}$, we compute the amplitude a GW background signal in our data would contribute to the timing residuals. By repeating this process while fitting only for spin frequency and spin frequency derivative, we calculate the total amount of GW background signal effectively in our data. Any timing fit will include spin frequency and spin frequency derivative, so the signal lost due to fitting those two parameters is effectively not in our data. This is shown in Equation 1, where $n$ is the number of TOAs, $\lambda_i^{\text{free}}$ are the eigenvalues of {\bf W}{\bf C}$_{{\bf r}}^{{\bf GW}}${\bf W} where all timing parameters were free, and $\lambda_i^{\text{fixed}}$ are the eigenvalues of {\bf W}{\bf C}$_{{\bf r}}^{{\bf GW}}${\bf W} where all the timing parameters were fixed except the spin frequency and spin frequency derivative. $\%GWs$ represents the percentage of assumed GWs present in the data after the fit.
\begin{equation}
\%GWs=\frac{\sum_{i=1}^{n} \sqrt{\lambda_i^{\text{free}}}}{\sum_{i=1}^{n} \sqrt{\lambda_i^{\text{fixed}}}} \times 100\%
\end{equation}
\subsection{Sensitivity vs.~Orbital Period}
In order to investigate how sensitivity to GWs changes with orbital period, using the \texttt{TEMPO2} plugin \texttt{FAKE} \citep{TEMPO2}, we simulated ten data sets with timing variations of rms 100 ns with no orbital frequency derivatives from PSR J0023+0923. It should be noted that PSR J0023+0293 shows very little orbital period variation and is currently being timed by NANOGrav \citep{NANOGrav9year,Robertsetal2014}. Each data set ranged from 51000 MJD to 54000 MJD. We repeated this process, changing the orbital period within the range 2.7 hours to 300 years. We then fit each set of simulated TOAs to its respective model with the correct orbital period. For each fit, we calculated the percent of a GW background signal left in the data after the fit and took the average value of the ten data sets with the same orbital period as the correct value. We then repeated this process fitting up to ten orbital frequency derivatives. The results are shown in Figure 1.
\subsection{Sensitivity vs.~Number of Modeled Orbital Frequency Derivatives}
\par
We used simulated data to quantify the effect on GW sensitivity of modeling many orbital frequency derivatives. We simulated five data sets from PSR J1748-2446ad, PSR J1748-2021D, PSR J1748-2446P, and PSR J2129-04 with rms 10$^{-4}$ from MJD 51000 to 54000. We chose these pulsars because each of them has a timing solution and requires modeling of several orbital frequency derivative terms. For each pulsar, each data set has a different number of orbital frequency derivative signals, ranging from no derivatives to four derivatives. We then made 17 model files for each data set, each modeling the simulated data with a different number of orbital frequency derivatives ranging from 0 to 16 derivatives. This may sometimes over-fit our simulated observations. We fit each model to each data set and computed the amount of GW background signal in the data after the fit for each fit. The results are shown in Figure 2.
\subsection{Transmission Spectrum of PSR J1748-2446P}
\par
In order to further demonstrate that the signal removed from modeling orbital frequency derivatives looks nothing like the signal of the GWs PTAs might see, we calculate the transmission spectrum of GWs for a timing model of based off PSR J1748-2446P, but with up to four OFDs modeled. First, we simulated pulse TOAs at different dates ranging from 51000 MJD to 54000 MJD, such that on average there is an observation every two weeks with white timing noise of rms 100 ns with up to four OFD signals in the data. Then, for each of the four timing models of PSR J1748-2446P with between one and four OFDs modeled, we fit the simulated TOAs to each corresponding timing model with the correct number of OFDs modeled. We then made a normalized Lomb-Scargle periodogram from the residuals of each timing fit. This process was repeated for 100 different sets of simulated TOAs in order to minimize noise. We plot the average Lomb-Scargle periodogram for each number of OFDs fit in Figure 3. This should correspond to the transmission spectrum described in \citet{Blandfordetal1984}.
%
\section{Results}
\begin{table*}[ht] 
\centering 
\begin{tabular}{p{2.5cm} p{1 cm}p{1 cm} p{1.1 cm}p{2cm} p{2cm} p{1.6cm} p{2cm} p{2cm}l c c c c c c c} 
\toprule 
 & & & & \multicolumn{2}{c}{\%GWs in Residuals} &\\
\cline{5-6}
\smallskip Pulsar &  Spin Period (ms)&  Orbital Period (hrs) &  \smallskip Observed OFDs &\smallskip \pbox{20cm}{0 OFD signals \\ 0 OFDs fit} &\smallskip \pbox{20cm}{4 OFD signals \\16 OFDs fit} & \smallskip \% Change \\ 
\midrule 
PSR J1748-2021D &13.50&6.87&2& 96.6 & 93.9 & 2.7 \\ 
PSR J1748-2446ad &1.40&26.23&6& 97.9 & 94.6 & 3.4 \\ 
PSR J1748-2446P &1.73&8.70&11& 97.9 & 94.9 & 3.1 \\ 
PSR J2129-04  &7.61&15.25&3 & 99.8 & 94.1 & 5.7 \\ 
\midrule 
\midrule 
Average &  & & & &  & 3.7 \\ 
\bottomrule 
\end{tabular}
\caption{The percent change in the amount of GWs that survive the fit after adding four orbital frequency derivative signals and fitting out 16 orbital frequency derivative terms.} 
\label{tab:template} 
\end{table*}
\subsection{Sensitivity vs.~Orbital Period}
\begin{figure}[b]
\centering
\includegraphics[width=0.5\textwidth]{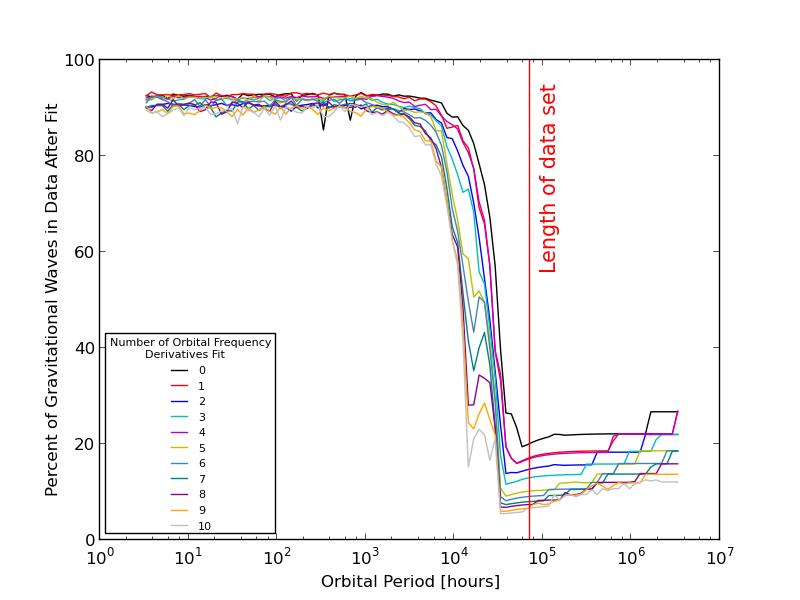}
\caption{Amount of a GW background signal remaining post-fit vs.\ orbital period for different numbers of orbital frequency derivative terms fit. We used simulated data from a timing solution for PSR J0023+0293 without orbital frequency derivative terms.}
\end{figure}
\par
We found that there are three GW sensitivity regimes in orbital period: a high sensitivity regime, a low sensitivity regime, and a transition regime. The low sensitivity regime covers orbital periods greater than the length of the data set, the transition regime covers orbital periods slightly less than the length of the data set, and the high sensitivity regime includes all orbital periods significantly shorter than the data span. This is expected as when the orbital period is greater than the length of the timing data set, GWs and orbital parameters are completely covariant. 
\par 
As shown in Figure 1, in the low sensitivity regime at the length of the data set, there is rougly a 65-75\% decrease in sensitivity between modeling no orbital frequency derivatives and modeling 10 orbital frequency derivatives. In the transition regime, there is roughly a 50-60\% decrease in sensitivity between modeling no orbital frequency derivatives and modeling 10 orbital frequency derivatives. In the high sensitivity regime, there is at most a 10\% decrease in sensitivity between modeling no orbital frequency derivatives and modeling 10 orbital frequency derivatives.
\subsection{Sensitivity vs.~Number of Modeled Orbital Frequency Derivatives}
\begin{figure}[t]
\centering
\includegraphics[width=0.5\textwidth]{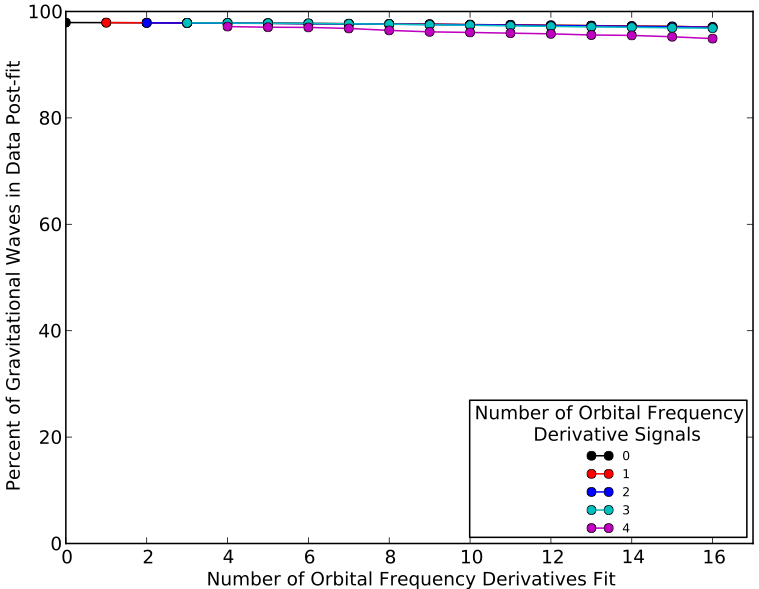}
\caption{Amount of a GW background signal not absorbed post-fit vs.\ number of orbital frequency derivatives fit for data sets with different numbers of orbital frequency derivative signals. We used simulated data from a timing solution of PSR J1748-2446ad (orbital period = 26.23 hrs) with the respective number of orbital frequency derivative signals.}
\end{figure}
\par
Figure 2 shows that modeling orbital frequency derivatives has very little effect on sensitivity to GWs and that the decrease in sensitivity due to modeling orbital frequency derivatives does not strongly depend on how many orbital frequency derivative signals are in the data. For PSR J1748-2446ad, there is a decrease in GW sensitivity of only 3.4\% between fitting no orbital frequency derivatives to a data set with no orbital frequency derivatives and fitting 16 orbital frequency derivatives to a data set with four orbital frequency derivatives. Furthermore, when this analysis was repeated for other pulsars, the decrease in GW sensitivity varied only slightly, giving an average decrease of 3.7\%. None of the pulsars showed a significant decrease in GW sensitivity due to modeling more orbital frequency derivatives.
\begin{figure*}
\centering
\includegraphics[width=1.\textwidth]{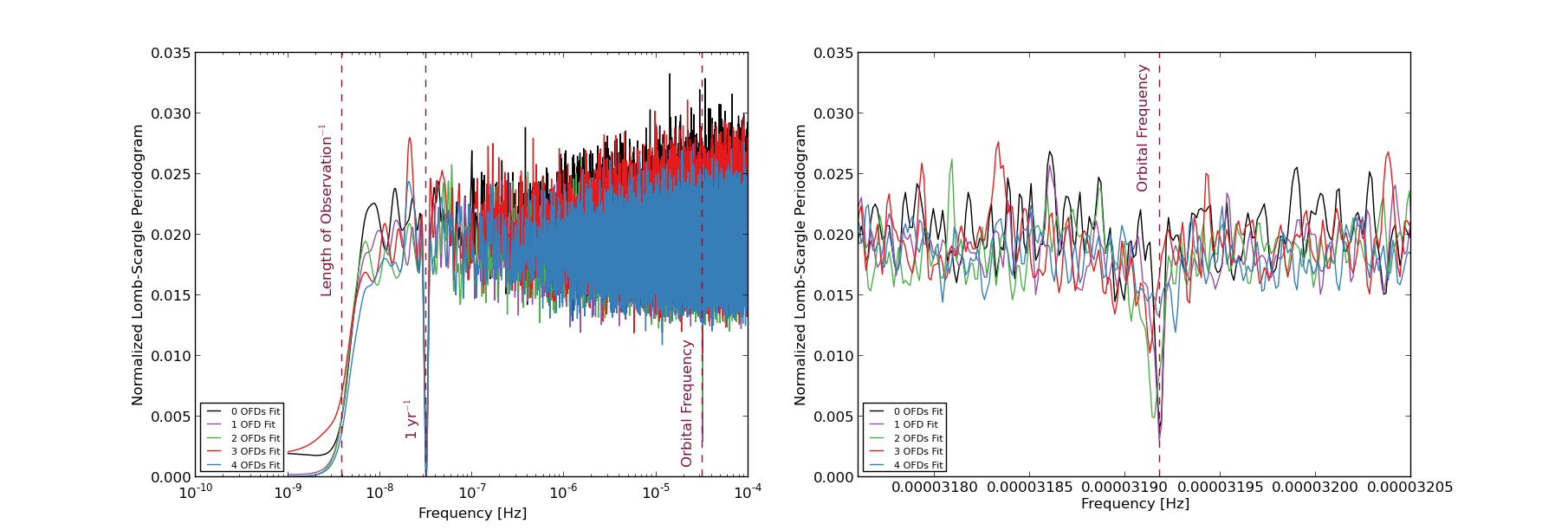}
\caption{GW Transmission spectrum for PSR J1748-2446P. Both panels show the normalized Lomb-Scargle periodogram of the timing fit residuals for simulated data from PSR J1748-2446P with up to four OFD signals. The left panel shows the entire spectrum while the right panel is zoomed in on the orbital frequency. Each color corresponds to the number of OFDs fit for in the model. The dashed vertical lines correspond to the inverse of the length of the data set, the frequency corresponding to a period of a year, and the orbital frequency. It is clear that as more OFDs are fit, the dip in Fourier power around the orbital frequency is spread out around the orbital frequency.}
\end{figure*}
\section{Discussion}
\subsection{Sensitivity vs.~Orbital Frequency}
\par
We expect to lose sensitivity to GWs when the orbital period is greater than the length of the data set. Because PTAs are most sensitive to frequencies on the order of $\frac{1}{T}$ \citep{Ellisetal2012}, where $T$ is the length of the data set, the full period of any GW detected will not be in the data. Therefore, when we search our data for GWs, we are searching in part for a signal that is an incomplete sine wave. As the length of the orbit approaches the length of the data set, the situation is analogous to the dramatic decrease in GW sensitivity when fitting for many spin period derivatives. The timing data set will not contain a full orbit, meaning the signal due to the orbit will be indistinguishable from the signal due to GWs. Thus, the GW signal and the signal from the orbit become covariant. In other words, any GW signal in our data will appear to be an error in orbital frequency, meaning the fit will absorb the timing variations due to GWs into the orbital frequency parameter, greatly reducing sensitivity to GWs. However, orbital periods of BWPs range from 2-20 hours and our timing data sets cover many years \citep{Chenetal2013}. Therefore, we only need to be concerned about the high sensitivity regime of orbital period, which shows modeling many orbital frequency derivatives has little effect on sensitivity to GWs.
\subsection{Sensitivity vs. Number of Modeled Orbital Frequency Derivatives}
\par
We should not be surprised that modeling orbital frequency derivatives does not significantly reduce sensitivity to GWs. The signal due to a binary orbit is a sine wave. The Fourier transform of a sine wave is a sharp spike at the frequency of the sine wave. Even when the frequency is slowly changing, as in the orbital frequency derivatives, all of the Fourier power remains around the orbital frequency. It takes less than a day for the a BWP system to complete an orbit, while it takes many years for a full GW to pass through the solar system, translating to a difference of more than three orders of magnitude in period. In other words, because the orbital timescales of BWPs and the timescales of GWs are dramatically different, the signal due to a binary orbit with changing frequency is very different from the signal due to GWs that PTAs are attempting to detect. 
\par
We can see this effect in Figure 3, which corresponds to the transmission spectrum of GWs for simulated TOAs from a modified model of PSR J1748-2446P. In this figure, we see both the transmission spectrum over a wide range of frequencies (left panel) and over a narrow window around the orbital frequency (right panel). The left panel shows three large decreases in GW transmission: one at frequencies lower than $\frac{1}{T}$, one at 1 year$^{-1}$, and one at the orbital frequency. The decrease in GW transmission at frequencies  lower than $\frac{1}{T}$ is expected since PTAs are not sensitive to GWs with periods longer than the length of the data set. Any GW of with a period of one year will be modeled as an error in the pulsar's position, leading to the transmission dip at 1 year$^{-1}$. Similarly, any GW of period equal to the orbital period will be modeled as an error in orbital period, producing the third decrease in GW transmission at the orbital frequency. The left panel also shows that the orbital period is orders of magnitude in frequency away from frequencies of detectable GWs. Furthermore, the right panel shows that as more OFDs are modeled, all the Fourier power removed is still around the orbital frequency and is merely broadened around the orbital frequency. Thus, the signals due to OFDs should not be covariant with detectable GWs. Because of this, we should not expect a large loss in GW sensitivity when accounting for more orbital frequency derivatives. Therefore, if the timing noise present in BWPs can be accounted for by modeling orbital frequency derivatives, they can be used in PTAs.
\subsection{Previous Results and Limitations}
Even though modeling timing noise as orbital frequency derivatives does not remove much sensitivity to gravitational waves, the long term timing results of the original BWP showed significant timing noise \citep{Niceetal2000}.  The largest amplitudes of the timing residuals were 30 $\mu$s at 430 MHz and 20 $\mu$s at 575 MHz \citep{Niceetal2000}. For this timing noise, \citet{Niceetal2000} offers the explanation of dispersion measure (DM) variations, but suggest that the timing noise lies within the system itself. We submit that this timing noise is reasonably explained solely by DM variations. Using the equation $t=\frac{e^2}{2\pi m_ec}\frac{\mathrm{DM}}{f^2}$ for the timing delay due to dispersion, where $t$ is the timing delay and $f$ the observing frequency \citep{LorimerandKramer}, we calculate that in order to explain the residuals of 30 $\mu$s at 430 MHz and 20 $\mu$s at 575 MHz, $\Delta$DMs of -0.0013 pc/cm$^3$ and 0.0016 pc/cm$^3$ are required, respectively. Furthermore, accounting for the timing noise between these two residuals requires the most extreme rate of change in DM. The necessary $\Delta$DM is 0.0029 pc/cm$^3$ over a period of two 
and a half years, requiring a rate of change in DM of 0.0012 pc/cm$^3$/yr. This value is well within the range of published DM variations for PTA MSPs \citep{NANOGrav9year,Keithetal2013}.
\section{Conclusions}
Given that the observed long term timing variations in PSR B1957+20 can reasonably be explained by DM variations, we recommend measuring the DM variations in this BWP in order to confirm our hypothesis. Furthermore, in order to test whether or not the timing variations seen by \citet{Niceetal2000} were due to DM variations, we recommend  long term multi-frequency timing observations of suitable BWPs. If the observed timing noise in black widow pulsars can be accounted for with DM variations and orbital frequency derivatives, we recommend using suitable BWPs (i.e.~non-eclipsing strong, fast, and narrow pulsars) in PTAs.
\par
BWPs are difficult to time accurately due to their orbital dynamics. However, this can be modeled effectively by orbital frequency derivatives. Even though we expect some GW sensitivity to be lost due to fitting for orbital frequency derivatives, we have shown that only a small amount of GW sensitivity is lost. 

\end{document}